\documentclass
[a4paper,pra,hyperref,superscriptaddress,showpacs,twocolumn]{revtex4}%
\usepackage{amsfonts}
\usepackage{amsmath}
\usepackage{amssymb}
\usepackage{graphicx}%
\providecommand{\U}[1]{\protect\rule{.1in}{.1in}}
\begin{document}
\title{Quantum and classical thermal correlations in the XY spin-1/2 chain}
\author{J. Maziero}
\affiliation{Centro de Ci\^{e}ncias Naturais e Humanas, Universidade Federal do ABC, R.
Santa Ad\'{e}lia 166, 09210-170, Santo Andr\'{e}, S\~{a}o Paulo, Brazil.}
\author{H. C. Guzman}
\affiliation{Centro de Ci\^{e}ncias Naturais e Humanas, Universidade Federal do ABC, R.
Santa Ad\'{e}lia 166, 09210-170, Santo Andr\'{e}, S\~{a}o Paulo, Brazil.}
\author{L. C. C\'{e}leri}
\affiliation{Centro de Ci\^{e}ncias Naturais e Humanas, Universidade Federal do ABC, R.
Santa Ad\'{e}lia 166, 09210-170, Santo Andr\'{e}, S\~{a}o Paulo, Brazil.}
\author{M. S. Sarandy}
\affiliation{Instituto de F\'{\i}sica, Universidade Federal Fluminense, Av. Gal. Milton
Tavares de Souza s/n, Gragoat\'{a}, 24210-346, Niter\'{o}i, Rio de Janeiro, Brazil.}
\author{R. M. Serra}
\affiliation{Centro de Ci\^{e}ncias Naturais e Humanas, Universidade Federal do ABC, R.
Santa Ad\'{e}lia 166, 09210-170, Santo Andr\'{e}, S\~{a}o Paulo, Brazil.}

\begin{abstract}
We investigate pairwise quantum correlation as measured by the quantum discord
as well as its classical counterpart in the thermodynamic limit of anisotropic
XY spin-1/2 chains in a transverse magnetic field for both zero and finite
temperatures. Analytical expressions for both classical and quantum
correlations are obtained for spin pairs at any distance. In the case of zero
temperature, it is shown that the quantum discord for spin pairs farther than
second-neighbors is able to characterize a quantum phase transition, even
though pairwise entanglement is absent for such distances. For finite
temperatures, we show that quantum correlations can be increased with
temperature in the presence of a magnetic field. Moreover, in the XX limit,
the thermal quantum discord is found to be dominant over classical correlation
while the opposite scenario takes place for the transverse field Ising model limit.

\end{abstract}

\maketitle

\section{Introduction}

Since its conception, quantum mechanics has been extensively applied to
condensed matter systems~\cite{CondMat}. This endeavour resulted in important
conceptual and technological advances. In the last few decades, this scenario
has gained even more strength with the born of quantum information science
(QIS)~\cite{QInf}. In this context, quantum spin systems have played a central
role in several applications for both quantum communication~\cite{Bose:03} and
quantum computation~\cite{Kane:98}. Indeed, quantum spins models describe the
effective interactions in a variety of physical systems~\cite{magnetism},
e.g., quantum Hall systems, high-temperature superconductors, heavy fermions,
and magnetic compounds. More generally, low dimensional systems as those
described by interacting spins are particularly interesting because of the
presence of typically pronounced quantum fluctuations as well as the
possibility of their realization by several distinct physical
approaches~\cite{Duan:03,Nori 2009}.

A key concept in QIS is the quantum correlation among parts of a composite
quantum system, which is a fundamental resource for several applications in
quantum information~\cite{QInf}. The existence of quantum correlations were
first noted in nonseparable (i.e., entangled) states. Entanglement was pointed
out by Schr\"{o}dinger\ in 1935~\cite{Schrodinger35} as the characteristic
trait of quantum mechanics. Since then, entanglement was basically the only
kind of quantum correlation theoretically and experimentally explored.
However, in the past few years, it has been realized that there exist
nonclassical correlations which are not captured by entanglement
measures~\cite{OllZur, OppenheimEtAl02}. In general, a quantum correlation,
which can be measured by the \textit{quantum discord} (QD)~\cite{OllZur},
often arises as a consequence of coherence between different partitions in a
quantum system, being present even in separable states. Recently, QD was
analyzed in a number of contexts, e.g., low dimensional spin
models~\cite{Dillenschneider08, Sarandy09,WerlangRigolin09,ChenLi09,ChenYin10}%
, open quantum systems~\cite{MazieroEtAl09,
ShabaniLidar09,Discord,FerraroEtAl09}, biological~\cite{Bradler:09}, and
relativistic~\cite{Rel} systems. Moreover, there exist strong indications that
the QD is the resource responsible for the speed up in the model of
computation known as deterministic quantum computation with one quantum
bit~\cite{Caves, White}.

In this article we consider pairwise QD in an infinite anisotropic XY spin-1/2
chain in the presence of an external transverse magnetic field for both zero
and finite temperatures. Our aim is to explore, in the thermodynamical regime,
the behavior of QD for spin pairs arbitrarily distant and also take into
account the effect of temperature on the behavior of correlations. Such
contributions, which have not been considered in previous works, will be shown
to bring several new effects to the subject. In particular, as we will show,
the QD for spin pairs more distant than second-neighbors is able to
characterize a quantum phase transition (QPT). This is a remarkable behavior,
since a signature of the QPT is then available even for distances where
pairwise entanglement is absent. Moreover, we will show that the QD may
increase with both temperature and magnetic field for certain regions of
parameter space. This result will extend, to the thermodynamic limit, the
previous analysis for two-spin Hamiltonians reported in
Ref.~\cite{WerlangRigolin09}. Finally, we will discuss the dominance of
quantum correlation over classical correlation for different limits of the XY
model, showing that the QD is greater than its classical counterpart for the
isotropic limit (XX model), with the opposite scenario taking place for the
transverse field Ising model. These results generalize those of
Refs.~\cite{Dillenschneider08} and \cite{Sarandy09}.

\section{Correlations in bipartite quantum systems}

The information-theoretical measure of the total correlation between the
partitions of a bipartite quantum state $\rho_{AB}$ is the quantum mutual
information~\cite{GroPoWi, SchuWest}
\begin{equation}
I\left(  \rho_{AB}\right)  =S\left(  \rho_{A}\right)  +S\left(  \rho
_{B}\right)  -S\left(  \rho_{AB}\right)  ,
\end{equation}
where $S\left(  \rho\right)  =-$Tr$\rho\log_{2}\rho$\ is the von Neumann
entropy and $\rho_{A}$ ($\rho_{B}$) is the reduced-density operator of the
partition $A$ ($B$). The nonclassical correlation present in $\rho_{AB}$ can
be quantified by the quantum discord, which is defined as~\cite{OllZur}
\begin{equation}
D\left(  \rho_{AB}\right)  \equiv I\left(  \rho_{AB}\right)  -C\left(
\rho_{AB}\right)  \text{,}\label{QC}%
\end{equation}
where
\begin{equation}
C\left(  \rho_{AB}\right)  \equiv S\left(  \rho_{A}\right)  -\min_{\left\{
\Pi_{j}\right\}  }S_{\left\{  \Pi_{j}\right\}  }\left(  \rho_{A\left\vert
B\right.  }\right)  \label{CC}%
\end{equation}
is the classical correlation present in the composite state $\rho_{AB}$
\cite{HenVed}. In Eq.~(\ref{CC}), the conditional entropy $S_{\left\{  \Pi
_{j}\right\}  }\left(  \rho_{A\left\vert B\right.  }\right)  $ can be defined
as
\begin{equation}
S_{\left\{  \Pi_{j}\right\}  }\left(  \rho_{A\left\vert B\right.  }\right)
=\sum_{j}q_{j}S(\rho_{A}^{j}),
\end{equation}
with $q_{j}=\operatorname*{Tr}\left[  \left(  \mathbf{1}_{A}\otimes\Pi
_{j}\right)  \rho_{AB}\left(  \mathbf{1}_{A}\otimes\Pi_{j}\right)  \right]  $
and $\rho_{A}^{j}=\left.  \operatorname*{Tr}\nolimits_{B}\left[  \left(
\mathbf{1}_{A}\otimes\Pi_{j}\right)  \rho_{AB}\left(  \mathbf{1}_{A}\otimes
\Pi_{j}\right)  \right]  \right/  q_{j}$. The minimum in Eq.~(\ref{CC}) is
taken over a complete set of projective measures $\left\{  \Pi_{j}\right\}  $
on the partition $B$. For pure states, we have that both quantum and classical
correlations are equal to entanglement entropy~\cite{GroPoWi, HenVed}. On the
other hand, for mixed states, entanglement is only a part of this nonclassical
correlation~\cite{OllZur, White, Caves}.

In order to compare our results for the QD with those for pairwise
entanglement, we will use the entanglement of formation ($\operatorname*{EoF}%
$) as a measure of entanglement. The concurrence ($\operatorname*{conc}$) is
monotonically related to the entanglement of formation by the following
expression
\[
\operatorname*{EoF}(\rho_{AB})=\mathcal{H}_{bin}\left\{  \left.  \left[
1+\sqrt{1-\left(  \operatorname*{conc}(\rho_{AB})\right)  ^{2}}\right]
\right/  2\right\}  ,
\]
where $\mathcal{H}_{bin}(x)=-x\log_{2}x-(1-x)\log_{2}(1-x)$ is the binary
entropy. For two qubits the concurrence reduces to the simple
form~\cite{Wootters98}
\begin{equation}
\operatorname*{conc}\left(  \rho_{AB}\right)  =\max\left(  0,\sqrt{\lambda
_{1}}-\sqrt{\lambda_{2}}-\sqrt{\lambda_{3}}-\sqrt{\lambda_{4}}\right)
\text{,}\label{Con}%
\end{equation}
where $\lambda_{i}$ ($i=1,2,3,4$) are the eigenvalues of $\rho_{AB}\tilde
{\rho}_{AB}$ in decreasing order, $\tilde{\rho}_{AB}=\left(  \sigma_{A}%
^{y}\otimes\sigma_{B}^{y}\right)  \rho_{AB}^{\ast}\left(  \sigma_{A}%
^{y}\otimes\sigma_{B}^{y}\right)  ^{\dagger}$, with $\rho_{AB}^{\ast}$ being
the conjugate of $\rho_{AB}$ in any basis, and $\sigma_{\alpha}^{y}$\ is the
$y$-component of the spin-1/2 Pauli operator for the partition $\alpha$
($\alpha=A,B$). We note that all the correlations quantifiers aforementioned
are measured in bits, as usual.

\section{Thermal correlations in the anisotropic XY spin chain}

The 1D XY model in a transverse field describes a chain of spins
anisotropically interacting in the $xy$ spin plane under the effect of a
magnetic field in the $z$ direction. The system is governed by the
Hamiltonian
\begin{equation}
H=-\sum_{j=0}^{N-1}\left\{  \frac{\lambda}{2}\left[  \left(  1+\gamma\right)
\sigma_{j}^{x}\sigma_{j+1}^{x}+\left(  1-\gamma\right)  \sigma_{j}^{y}%
\sigma_{j+1}^{y}\right]  +\sigma_{j}^{z}\right\}  \text{,}%
\end{equation}
where $\sigma_{j}^{k}$ $\left(  k=x,y,z\right)  $ is the $k$ component of the
spin-1/2 Pauli operator acting on site $j$ state space, $\gamma$ is the degree
of anisotropy (we take for simplicity $0\leq\gamma\leq1$), and $\lambda$
provides the strength of the inverse of the external transverse magnetic
field. We will be interested in the limit of an infinite chain, namely,
$N\rightarrow\infty$.

The XY model is exactly solvable~\cite{BarouchMcCoy70,BarouchMcCoy71}. The
Hamiltonian can be diagonalized via a Jordan-Wigner map followed by a
Bogoliubov transformation (see, e.g., Ref.~\cite{Chakrabarti96}). By taking
the thermal ground state, the reduced density operator for the sites $0$ and
$n$ reads~\cite{Nielsen02}
\begin{equation}
\rho_{0n}=\frac{1}{4}\left\{  I_{0n}+\langle\sigma^{z}\rangle\left(
\sigma_{0}^{z}+\sigma_{n}^{z}\right)  +\sum_{k=1}^{3}\langle\sigma_{0}%
^{k}\sigma_{n}^{k}\rangle\sigma_{0}^{k}\sigma_{n}^{k}\right\}  \text{,}
\label{TS}%
\end{equation}
with $I_{0n}$ being the identity operator acting on the state space of the
sites $0n$. Although an unbroken state, $\rho_{0n}$ is able to provide an
exact description of the critical behavior as well as its scaling in finite
systems~\cite{Nielsen02,Osterloh:02,Gu:03,Gu:04,Wu:04} (For a detailed
treatment of the spontaneous symmetry breaking at zero temperature see
Refs.~\cite{Syljuasen:03,Osterloh:06,Oliveira:08}).

Since the system is invariant by translations, the elements of the two site
reduced-density operator depends only on the distance ($n$) between the sites.
The transverse magnetization is given by~\cite{BarouchMcCoy70}
\begin{equation}
\langle\sigma^{z}\rangle=-\int_{0}^{\pi}\frac{\left(  1+\lambda\cos
\phi\right)  \tanh\left(  \beta\omega_{\phi}\right)  }{2\pi\omega_{\phi}}%
d\phi\text{,}%
\end{equation}
where $\omega_{\phi}=\sqrt{\left(  \gamma\lambda\sin\phi\right)  ^{2}+\left(
1+\lambda\cos\phi\right)  ^{2}}/2$ and $\beta=1/kT$ with $k$ being the
Boltzmann's constant and $T$ the absolute temperature. The two-point
correlation functions read~\cite{BarouchMcCoy71}
\begin{equation}
\langle\sigma_{0}^{x}\sigma_{n}^{x}\rangle=%
\begin{vmatrix}
G_{-1} & G_{-2} & \cdots & G_{-n}\\
G_{0} & G_{-1} & \cdots & G_{-n+1}\\
\vdots & \vdots & \ddots & \vdots\\
G_{n-2} & G_{n-3} & \cdots & G_{-1}%
\end{vmatrix}
\text{,}%
\end{equation}%
\begin{equation}
\langle\sigma_{0}^{y}\sigma_{n}^{y}\rangle=%
\begin{vmatrix}
G_{1} & G_{0} & \cdots & G_{-n+2}\\
G_{2} & G_{1} & \cdots & G_{-n+3}\\
\vdots & \vdots & \ddots & \vdots\\
G_{n} & G_{n-1} & \cdots & G_{1}%
\end{vmatrix}
\text{,}%
\end{equation}
and%
\begin{equation}
\langle\sigma_{0}^{z}\sigma_{n}^{z}\rangle=\langle\sigma^{z}\rangle^{2}%
-G_{n}G_{-n}\text{,}%
\end{equation}
where%
\begin{align}
G_{n}  &  =\int_{0}^{\pi}\frac{\tanh\left(  \beta\omega_{\phi}\right)  }%
{2\pi\omega_{\phi}}\left\{  \cos\left(  n\phi\right)  \left(  1+\lambda
\cos\phi\right)  \right. \nonumber\\
&  \left.  -\gamma\lambda\sin\left(  n\phi\right)  \sin\phi\right\}
d\phi\text{.}%
\end{align}

The total correlation in (\ref{TS}) is quantified by the quantum mutual
information as $I(\rho_{0n})=S(\rho_{0})+S(\rho_{n})-S(\rho_{0n})$ with
$S(\rho_{0})=S(\rho_{n})=-%
%TCIMACRO{\tsum \nolimits_{i=0}^{1}}%
%BeginExpansion
{\textstyle\sum\nolimits_{i=0}^{1}}
%EndExpansion
\{[1+(-1)^{i}\langle\sigma^{z}\rangle]/2\}\log_{2}\{[1+(-1)^{i}\langle
\sigma^{z}\rangle]/2\}$ and $S(\rho_{0n})=-%
%TCIMACRO{\tsum \nolimits_{i,j=0}^{1}}%
%BeginExpansion
{\textstyle\sum\nolimits_{i,j=0}^{1}}
%EndExpansion
(\xi_{i}\log_{2}\xi_{i}+\xi_{j}\log_{2}\xi_{j})$, where $\xi_{i}%
=[1+\langle\sigma_{0}^{z}\sigma_{n}^{z}\rangle+(-1)^{i}\sqrt{(\langle
\sigma_{0}^{x}\sigma_{n}^{x}\rangle-\langle\sigma_{0}^{y}\sigma_{n}^{y}%
\rangle)^{2}+4\langle\sigma^{z}\rangle^{2}}]/4$ and $\xi_{j}=[1-\langle
\sigma_{0}^{z}\sigma_{n}^{z}\rangle+(-1)^{j}(\langle\sigma_{0}^{x}\sigma
_{n}^{x}\rangle+\langle\sigma_{0}^{y}\sigma_{n}^{y}\rangle)]/4$. We can
compute the QD and its classical counterpart by extremizing Eqs.~(\ref{QC})
and~(\ref{CC}) over the following complete set of orthonormal projectors
$\left\{  \Pi_{\beta}=\left\vert \Theta_{\beta}\right\rangle \left\langle
\Theta_{\beta}\right\vert ,\beta=\parallel,\perp\right\}  $ onto the $n$th
nearest-neighbor, where $\left\vert \Theta_{\parallel}\right\rangle \equiv
\cos(\theta/2)\left\vert 0\right\rangle _{n}+e^{i\varphi}\sin(\theta
/2)\left\vert 1\right\rangle _{n}$ and $\left\vert \Theta_{\perp}\right\rangle
\equiv e^{-i\varphi}\sin(\theta/2)\left\vert 0\right\rangle _{n}-\cos
(\theta/2)\left\vert 1\right\rangle _{n}$. Remarkably, a numerical analysis
implies that the extremization is achieved, for any values of $\gamma$,
$\lambda$, and $T$, by the choice $\theta=\pi/2$ and $\varphi=0$
\cite{Comment}. Then, the relevant measurement for the model is given by
$\{|+\rangle\langle+|,|-\rangle\langle-|\}$, with $|+\rangle$ and $|-\rangle$
denoting the up and down spins in the $x$ direction, namely, $|\pm
\rangle=(|\uparrow\rangle^{z}\pm|\downarrow\rangle^{z})/\sqrt{2}$. This result
generalizes the extremization obtained for the transverse field Ising model at
zero temperature in Ref.~\cite{Sarandy09}, and allows us to write the
classical correlation as
\begin{equation}
C(\rho_{0n})=\mathcal{H}_{bin}\left(  p_{1}\right)  -\mathcal{H}_{bin}\left(
p_{2}\right)  ,\label{cc-final-ib}%
\end{equation}
where
\begin{subequations}
\begin{align}
p_{1} &  =\frac{1}{2}\left(  1+\langle\sigma^{z}\rangle\right)  ,\\
p_{2} &  =\frac{1}{2}\left(  1+\sqrt{\langle\sigma_{0}^{x}\sigma_{n}%
^{x}\rangle^{2}+\langle\sigma^{z}\rangle^{2}}\right)  .\label{p-ib}%
\end{align}
Thus the quantum correlation in state (\ref{TS}) is simply given by
\end{subequations}
\begin{equation}
D(\rho_{0n})=\mathbf{\ }I(\rho_{0n})-\mathbf{\ }C(\rho_{0n})\text{.}%
\end{equation}

\subsection{Correlations at zero temperature and QPTs}

Let us first consider the XY model at zero temperature. Such a model has a
quantum phase diagram displayed in Fig.~1 (see,
e.g.,~\cite{BarouchMcCoy70,BarouchMcCoy71,Latorre:04}).%

%TCIMACRO{\FRAME{fhFU}{2.2658in}{2.2987in}{0pt}{\Qcb{(Color online) Quantum
%phase diagram for the anisotropic XY spin 1/2 chain. The XX model obtained by
%setting $\gamma=0$, displays a critical line for $\lambda\in\left[
%0,1\right]  $. The Ising model obtained for $\gamma=1$, exhibits a critical
%point at $\lambda=1$. }}{\Qlb{fig0}}{fig1.eps}%
%width 2.2658in;  height 2.2987in;  depth 0pt;  original-width 5.6092in;
%original-height 5.6948in;  cropleft "0";  croptop "1";  cropright "1";
%cropbottom "0";  filename 'fig1.eps';file-properties "XNPEU";}}}%
%BeginExpansion
\begin{figure}
[h]
\begin{center}
\includegraphics[
height=2.2987in,
width=2.2658in
]%
{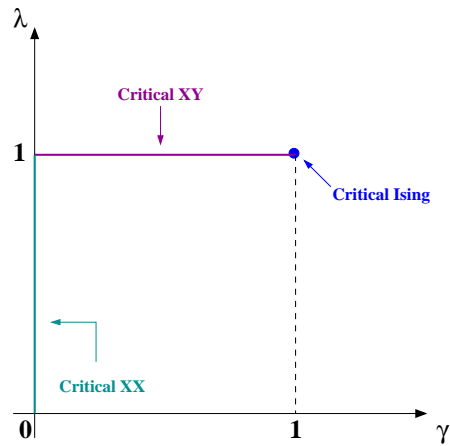}%
\caption{(Color online) Quantum phase diagram for the anisotropic XY spin 1/2
chain. The XX model obtained by setting $\gamma=0$, displays a critical line
for $\lambda\in\left[  0,1\right]  $. The Ising model obtained for $\gamma=1$,
exhibits a critical point at $\lambda=1$. }%
\label{fig0}%
\end{center}
\end{figure}
%EndExpansion
The QD for the first, second, third, and fourth nearest-neighbors in the
thermal ground state (\ref{TS}), close to zero temperature, is displayed in
Fig. 2 (a)-(d). As expected, QD decreases as we increase the distance between
the sites. However, we also see a clear difference in the amount of quantum
correlation between the regions where $\lambda<1$ and $\lambda>1$. Although
the maximum value of QD decreases as we increase the distance between the
sites, the slope in the critical region gets more evident for far neighbors.
The maximum increasing rate of the QD as function of $\lambda$ occurs at the
quantum phase transition line ($\lambda=1$). We also note that nonclassical
correlation is created when the magnetic field increases.%
%TCIMACRO{\FRAME{ftbpFU}{3.3762in}{4.4261in}{0pt}{\Qcb{Quantum discord between
%(a) first, (b) second, (c) third, and (d) fourth nearest-neighbors and EoF
%between (e) third and (f) fourth nearest-neighbors as a function of anisotropy
%($\gamma$) and $\lambda$\ at zero temperature.}}{\Qlb{fig1}}{fig2.eps}%
%width 3.3762in;  height 4.4261in;  depth 0pt;  original-width 7.6588in;
%original-height 10.0733in;  cropleft "0";  croptop "1";  cropright "1";
%cropbottom "0";  filename 'fig2.eps';file-properties "XNPEU";}}}%
%BeginExpansion
\begin{figure}
[ptb]
\begin{center}
\includegraphics[
height=4.4261in,
width=3.3762in
]%
{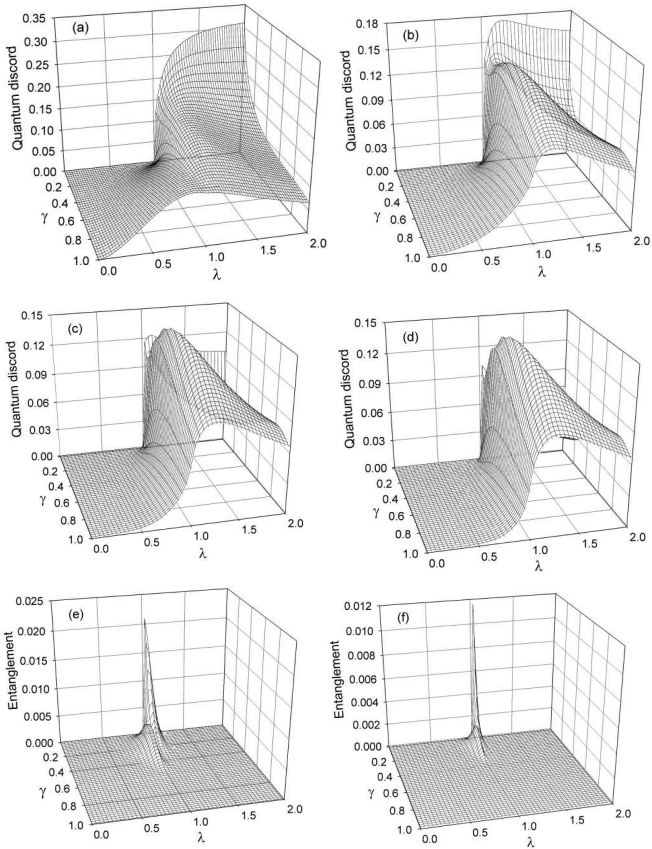}%
\caption{Quantum discord between (a) first, (b) second, (c) third, and (d)
fourth nearest-neighbors and EoF between (e) third and (f) fourth
nearest-neighbors as a function of anisotropy ($\gamma$) and $\lambda$\ at
zero temperature.}%
\label{fig1}%
\end{center}
\end{figure}
%EndExpansion
The derivative of the quantum discord between fourth nearest-neighbors with
respect to $\lambda$ is depicted in Fig.~3. So, the quantum discord between
far neighbors can be used to characterize the QPT. It is important to mention
that, even when entanglement is not present, the QPT can be clearly revealed
through the singular behavior of QD. In Fig. 2 (e) and (f) we note that, for a
considerable range of $\gamma$, EoF is zero (there is no entanglement) for
third and fourth nearest-neighbors. In the particular case of the transverse
field Ising model ($\gamma=1$), entanglement is indeed completely absent for
sites farther than second nearest-neighbors~\cite{Osterloh:02}. On the other
hand, QD is non-null\ (where the entanglement is zero) and its behavior
reveals that, for $\gamma=1$ and $\lambda>1$, we have long-range nonclassical
correlation between far sites. It is worthwhile to observe that the pairwise
classical correlation can also be employed to detect a QPT in a very evident
way. In the Ising model, for example, the first derivative of the classical
correlation with respect to $\lambda$ is not analytic at the critical point,
while the second derivative of quantum discord presents such a nonanalicity
\cite{Sarandy09}. This behavior of the classical correlation also holds in the
XY model for the whole range of values of $\gamma$ considered here, as
depicted in Fig. 4.%

%TCIMACRO{\FRAME{ftbpFU}{3.1868in}{2.6411in}{0pt}{\Qcb{(Color online)
%Derivative of the quantum discord between fourth nearest-neighbors with
%respect to $\lambda$ at zero temperature.}}{\Qlb{fig2}}{fig3.eps}%
%width 3.1868in;  height 2.6411in;  depth 0pt;  original-width 6.5683in;
%original-height 5.4353in;  cropleft "0";  croptop "1";  cropright "1";
%cropbottom "0";  filename 'fig3.eps';file-properties "XNPEU";}}}%
%BeginExpansion
\begin{figure}
[ptb]
\begin{center}
\includegraphics[
height=2.6411in,
width=3.1868in
]%
{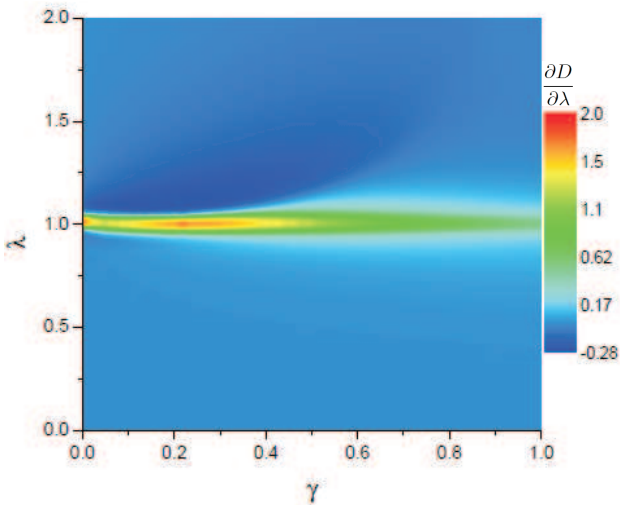}%
\caption{(Color online) Derivative of the quantum discord between fourth
nearest-neighbors with respect to $\lambda$ at zero temperature.}%
\label{fig2}%
\end{center}
\end{figure}
%EndExpansion%
%TCIMACRO{\FRAME{ftbpFU}{3.1401in}{2.5797in}{0pt}{\Qcb{(Color online)
%Derivative of the classical correlation between fourth nearest-neighbors with
%respect to $\lambda$ at zero temperature.}}{\Qlb{DCC}}{fig4.eps}%
%width 3.1401in;  height 2.5797in;  depth 0pt;  original-width 7.2376in;
%original-height 5.9404in;  cropleft "0";  croptop "1";  cropright "1";
%cropbottom "0";  filename 'fig4.eps';file-properties "XNPEU";}}}%
%BeginExpansion
\begin{figure}
[ptb]
\begin{center}
\includegraphics[
height=2.5797in,
width=3.1401in
]%
{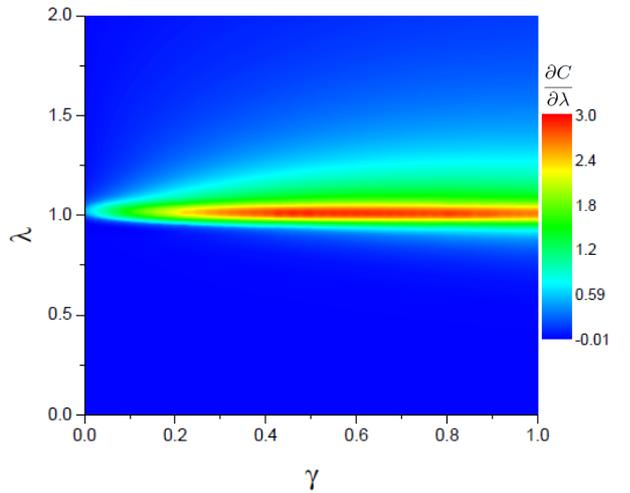}%
\caption{(Color online) Derivative of the classical correlation between fourth
nearest-neighbors with respect to $\lambda$ at zero temperature.}%
\label{DCC}%
\end{center}
\end{figure}
%EndExpansion

\subsection{Correlations at finite temperatures}

In order to introduce finite temperature, we begin by considering the thermal
state of the XX model ($\gamma=0)$. For this limit, we plot the thermal
classical and quantum correlations for the second nearest-neighbors in Fig. 5.
Similar results are obtained for others than second nearest-neighbors. We note
that the quantum discord is typically greater than its classical counterpart
and it indeed may increase with temperature (in a given region) for some
values of $\lambda$. It is evidenced near to the critical value $\lambda=1$.
This result extends, to the thermodynamic limit, the previous analysis for
two-spins Hamiltonians reported in Ref. \cite{WerlangRigolin09}. The
increasing of QD with temperature when a magnetic field is present is a
consequence of the fact that, when the field is turned on, the ground state
tends to be less correlated than some low-lying excited states. So the effect
of the temperature is to populate such correlated excited states, leading to
the net effect of an increasing of QD. Naturally, this effect tends to
disappear as the temperature gets too large. This is similar to the behavior
of entanglement observed in Ref.~\cite{Arnesen:01}.%
%TCIMACRO{\FRAME{fhFU}{3.4229in}{1.4209in}{0pt}{\Qcb{(a) Classical and \ (b)
%quantum correlations between second nearest-neighbors in the XX model as a
%function of temperature ($kT$) and inverse of the magnetic field ($\lambda$%
%).}}{\Qlb{XX}}{fig5.eps}{\special{ language "Scientific Word";
%type "GRAPHIC";  maintain-aspect-ratio TRUE;  display "USEDEF";
%valid_file "F";  width 3.4229in;  height 1.4209in;  depth 0pt;
%original-width 7.7366in;  original-height 3.1981in;  cropleft "0";
%filename 'fig5.eps';file-properties "XNPEU";}}}%
%BeginExpansion
\begin{figure}
[h]
\begin{center}
\includegraphics[
height=1.4209in,
width=3.4229in
]%
{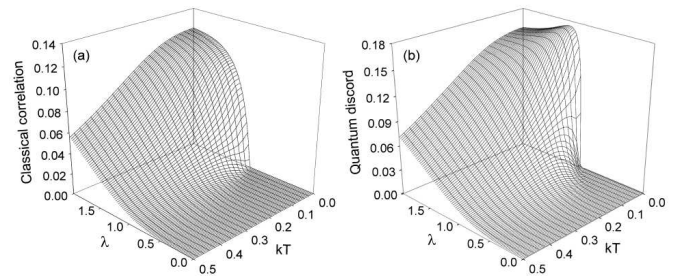}%
\caption{(a) Classical and \ (b) quantum correlations between second
nearest-neighbors in the XX model as a function of temperature ($kT$) and
inverse of the magnetic field ($\lambda$).}%
\label{XX}%
\end{center}
\end{figure}
%EndExpansion%
%TCIMACRO{\FRAME{ftFU}{3.4316in}{1.4364in}{0pt}{\Qcb{(a) Classical and
%(b)\ quantum correlations between second nearest-neighbors in the transverse
%Ising model as a function of temperature ($kT$) and inverse of the magnetic
%field ($\lambda$).}}{\Qlb{Ising}}{fig6.eps}%
%width 3.4316in;  height 1.4364in;  depth 0pt;  original-width 7.7366in;
%original-height 3.2232in;  cropleft "0";  croptop "1";  cropright "1";
%cropbottom "0";  filename 'fig6.eps';file-properties "XNPEU";}}}%
%BeginExpansion
\begin{figure}
[t]
\begin{center}
\includegraphics[
height=1.4364in,
width=3.4316in
]%
{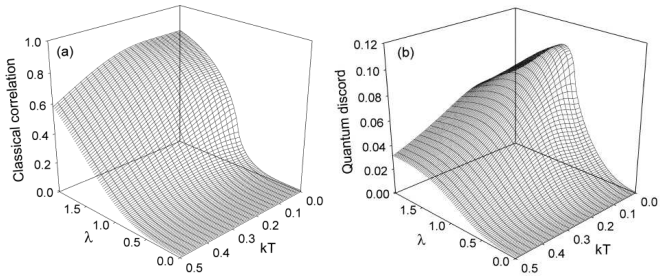}%
\caption{(a) Classical and (b)\ quantum correlations between second
nearest-neighbors in the transverse Ising model as a function of temperature
($kT$) and inverse of the magnetic field ($\lambda$).}%
\label{Ising}%
\end{center}
\end{figure}
%EndExpansion

Let us now turn our attention to the transverse Ising model ($\gamma=1$) at
finite temperatures. Classical correlation and QD for second nearest-neighbors
are shown in Fig.~6. Note that QD still increases with temperature but this
effect is feeble when compared with the same behavior in XX model. We also
observe that the classical correlation is typically greater than the quantum
discord for any temperature. This is the opposite scenario in comparison with
the XX model. Remarkably, for a weak magnetic field, QD available in the XX
model overcomes that of the transverse Ising model. So, applications of the XY
model in QIS tends to offer more quantum correlation as a resource in the
isotropic limit (the XX model) than in the Ising limit.

\section{Conclusion}

%%%% v2-revised
In summary, we have examined pairwise QD and its classical counterpart in the
thermodynamic limit of the anisotropic XY spin-1/2 chain in the presence of an
external transverse magnetic field. We have considered the system at both zero
and finite temperatures, providing an analytical expression for both classical
and quantum correlations for spin pairs at any distance. Remarkably, we have
shown that the quantum discord between far neighbors is able to characterize a
QPT, even for distances where pairwise entanglement is absent. This is a
consequence of the longer range of nonclassical correlation (that is
nonvanishing for a class of separable states) in comparison with the
short-range behavior of pairwise entanglement. Concerning the thermal effect
onto correlations, we have shown how QD can be increased with temperature as
the transverse magnetic field is varied. Moreover, we have also outlined the
dominance of the QD over classical correlation for the XX model in opposition
to the Ising limit. Generalization of these results for larger subsystems and
further analysis of the long-range behavior of QD for quantum critical
phenomena including spontaneous symmetry breaking (at zero temperature) are
left for a future investigation.
%%%%%%%

\begin{acknowledgments}
We are grateful for the financial support from CNPq, CAPES, FAPESP, FAPERJ,
and UFABC. This work was performed as part of the Brazilian National Institute
for Science and Technology of Quantum Information (INCT-IQ).
\end{acknowledgments}


\begin{thebibliography}{99}
\bibitem {Comment}After the completion of this work we became aware of an
analytical expression (see Ref.~\cite{Ali}) for the classical correlation in a
class of states that includes the $Z_{2}$-symmetric states. This analytical
result confirms our numerical verification.
\end{thebibliography}
\end{document}